\documentclass[12pt,a4paper]{iopart}
\usepackage{iopams}
\usepackage{graphicx}
\usepackage[usenames]{color}
\usepackage{ulem}

\begin{document}

\title[]{On the characterization of breather and rogue wave solutions of an inhomogeneous nonlinear Schr\"{o}dinger equation}
\author{K. Manikandan and M. Senthilvelan}
\address{Centre for Nonlinear Dynamics, School of Physics,  Bharathidasan University, \\Tiruchirappalli - 620 024, Tamil Nadu, India}

\begin{abstract}
We construct breather and rogue wave solutions of a variable coefficient nonlinear Schr\"{o}dinger equation with an external linear potential. This generalized model describes the nonlinear wave propagation in an inhomogeneous plasma/medium.  We derive several localized solutions including Ma breather, Akhmediev breather, two-breather and rogue wave solutions of this model and show how the inhomogeneity of space modifies the shape and orientation of these localized structures.  We also depict the trajectories of the inhomogeneous rogue wave. Our results may be useful for controlling plasmonic energy along the plasma surface.
\end{abstract}

\section{Introduction}
The study of nonlinear wave propagation in dispersive and inhomogeneous media is of great interest.  It has a wide range of applications such as radio waves in the ionosphere \cite{budd}, waves in the ocean \cite{osbore}, optical pulses in glass fibers \cite{hase}, laser radiation in plasma \cite{tewari} and so on.  The propagation of a general dimensionless nonlinear wave envelope in a weakly inhomogeneous plasma/medium obeys the following variable coefficients nonlinear Schr\"{o}dinger equation (NLS) equation \cite{chen:liu,bala,serkin}, namely
\begin{equation}
i\psi_t+\frac{1}{2}\psi_{xx}+h(t)\vert \psi\vert ^2\psi+M(x,t)\psi=0, \label{wav:eqn}
\end{equation}
where $\psi(x,t)$ represents the complex envelope of the physical field, $x$ is the longitudinal variable, $t$ is the transverse variable and subscripts denote partial derivative with respect to that variable, $h(t)$ and $M(x,t)$ refer the nonlinearity and inhomogeneity management parameters respectively.  The study of inhomogeneous NLS (INLS) equation with longitudinally and transversely varying inhomogeneity is of contemporary interest in several branches of physics \cite{atre:pani,solli,yan,zhong}.  In this paper, we choose the inhomogeneity of the linear potential $M(x,t)$ to be $2kx$, so that Eq. (\ref{wav:eqn}) takes the following form \cite{chen:liu,bala}  
\begin{equation}
i\psi_t+\frac{1}{2}\psi_{xx}+h(t)\vert \psi\vert ^2\psi+2kx\psi=0, \label{aa1}
\end{equation}
where $k$ is the inhomogeneity parameter.  During the past two decades several investigations have been made on to study how the inhomogeneous medium affects the propagation of solitary waves \cite{chen:liu,bala,belm,haseg}.  However only fewer works have been devoted to analyze how the inhomogeneity of space and time affects the other localized solutions like breather and rogue waves (RWs) \cite{dai,yang,xfwu,wang,khaw,shuk}.  In this work we intend to concentrate on this particular aspect.

Breather is a localized solution with temporally or spatially periodic structure and appears to be the internal oscillations and bound states of nonlinear wave packets \cite{mande}.  A RW is a wave which is localized in both space and time and appears from nowhere and disappears without a trace \cite{karif}.  A wave is classified under this category when its wave height (distance from trough to crest) reaches a value which is at least twice the significant wave height \cite{kharif,osborn}.  Even though it was first observed in arbitrary depth of ocean, the phenomenon is now shown to appear in diverse areas of physics \cite{blud:kono,kibler,chab:hoff,bailung,shat,ya,mosl}.  Certain kinds of exact solutions of NLS equation such as Peregrine soliton \cite{pere,akmv:anki}, time periodic breather or Ma soliton (MS) \cite{ma} and space periodic breather or Akhmediev breather (AB) \cite{eleon,kadz} have been considered to describe the possible mechanism for the formation of RWs.

Motivated by the contemporary development in the investigation of the breather and RW solutions of NLS type equations, in this paper, we construct Ma breather, AB, two-breathers, first-, second-, and third-order RW solutions with and without free parameters through the similarity transformation method.  We also investigate in-detail the impact of inhomogeneity on these localized solutions.  To construct exact solutions of (\ref{aa1}) one may look for a transformation that can map the given equation to the standard NLS equation.  The most popular way of transforming INLS equation to the NLS equation is the similarity transformation method \cite{serkin, belm,susl,pono}.  The necessary similarity transformation can be obtained by considering a generalized transformation $\psi(x,t)$=$\frac{1}{\sqrt{\mu(t)}}\phi(\xi,\tau)\exp[i (\alpha(t)x^2+\delta(t)x+\kappa(t))]$, where $\xi=\beta(t)x+\epsilon(t)$, $\tau=\gamma(t)$ and $\mu(t)$, $\alpha(t)$, $\beta(t)$, $\gamma(t)$, $\delta(t)$, $\epsilon(t)$ and $\kappa(t)$ are arbitrary functions of $t$, which can map Eq. (\ref{aa1}) to the NLS equation 
\begin{equation}
i\phi_{\tau}+\frac{1}{2}\phi_{\xi\xi}+ \vert \phi\vert ^2\phi=0. \label{a3}
\end{equation}
Substituting this transformation in (\ref{aa1}) and eliminating $\phi_{\tau}$ by Eq. (\ref{a3}) one can obtain a set of differential equations for these unknown arbitrary functions.  Solving them we can find these arbitrary functions which in turn fix the exact form.  As a result   
\small
\begin{eqnarray}
\label{a16}
\psi(x,t)&=&\frac{\phi(\xi,\tau)}{\sqrt{\mu_0(1+2\alpha_0 t)}}\exp\left[i \left(\frac{\alpha_0}{1+2\alpha_0 t}x^2 +\left(kt+\frac{\delta_0+kt}{1+2\alpha_0 t}\right)x \nonumber  \right.\right. \\ && \left.\left.  +\kappa_0-\frac{k^2t^3}{6}-\frac{t(\delta_0+kt)^2}{2(1+2\alpha_0 t)}\right)\right],
\end{eqnarray}
\normalsize
where $\phi(\xi,\tau)$ is the solution of NLS equation and provided $h(t)$ to be $\displaystyle{\frac{h_0 \mu_0 \beta_0^2}{1+2\alpha_0 t}}$.  With this choice of $h(t)$ we have an integrable INLS equation of the form 
\begin{eqnarray}
\label{a17}
i\psi_t+\frac{1}{2}\psi_{xx}+\frac{\mu_0 \beta_0^2}{1+2\alpha_0 t}\vert \psi\vert ^2\psi+2kx\psi=0,
\end{eqnarray}
where we have taken $h_0=1$ without loss of generality.  In the above $h_0$, $\mu_0$, $\alpha_0$, $\beta_0$, $\gamma_0$, $\delta_0$, $\epsilon_0$ and $\kappa_0$ are integration constants.  Since (\ref{a17}) can be mapped to the standard NLS equation we can generate certain new localized structures including breather and rogue wave (RW) solutions and study how these localized solutions are affected by the inhomogeneity parameter.  

The paper is organized as follows.  In Sec. 2, we construct AB, Ma and two-breather solutions to INLS equation (\ref{a17}) and study their characteristics in detail.  In Sec. 3, we construct RW solutions without and with free parameters and investigate how these RW structures get modified in the plane wave background.  In Sec. 4, we examine certain characteristics of RW, namely the evolution of its peak and width and depict the trajectories of inhomogeneous RW.  Finally, in section 5, we present a summary of our results and conclusions.

\section{Characteristics of breathers}
To begin with we construct first- and second-order breather solutions of (\ref{a17}).  The breather solution of NLS equation is given by \cite{eleon}
\begin{equation}
\label{a18}
\phi_1(\xi,\tau)=\left[\frac{m^2 \cosh(d \tau_s) + 2 i m v \sinh(d \tau_s)}{2 (\cosh(d \tau_s) - v \cos(m (\xi - \xi_1)))}-1\right]e^{i\tau},
\end{equation}
where $\tau_s$=$\tau - \tau_1$, the parameters $m$ and $v$ are expressed in terms of a complex eigenvalue (say $l$), that is $m=2\sqrt{1+l^2}$ and $v$ = $Im(l$), and $\xi_1$ and $\tau_1$ serve as coordinate shifts from the origin.  The parameter $d(=m v)$ in (\ref{a18}) is the growth rate of modulation instability.  Substituting the breather solution (\ref{a18}) in (\ref{a16}) we can capture the breather solution of (\ref{a17}).  When $v$ lies between $0$ and $1$ and $m$ is real, one can obtain the AB solution which is periodic in $x$ and localized in $t$.  On the other hand when $v>1$ and $m$ is imaginary, Eq. (\ref{a16}) provides Ma breather solution which is periodic in $t$ and localized in $x$.  
\begin{figure}[!ht]
\begin{center}
\includegraphics[width=0.85\linewidth]{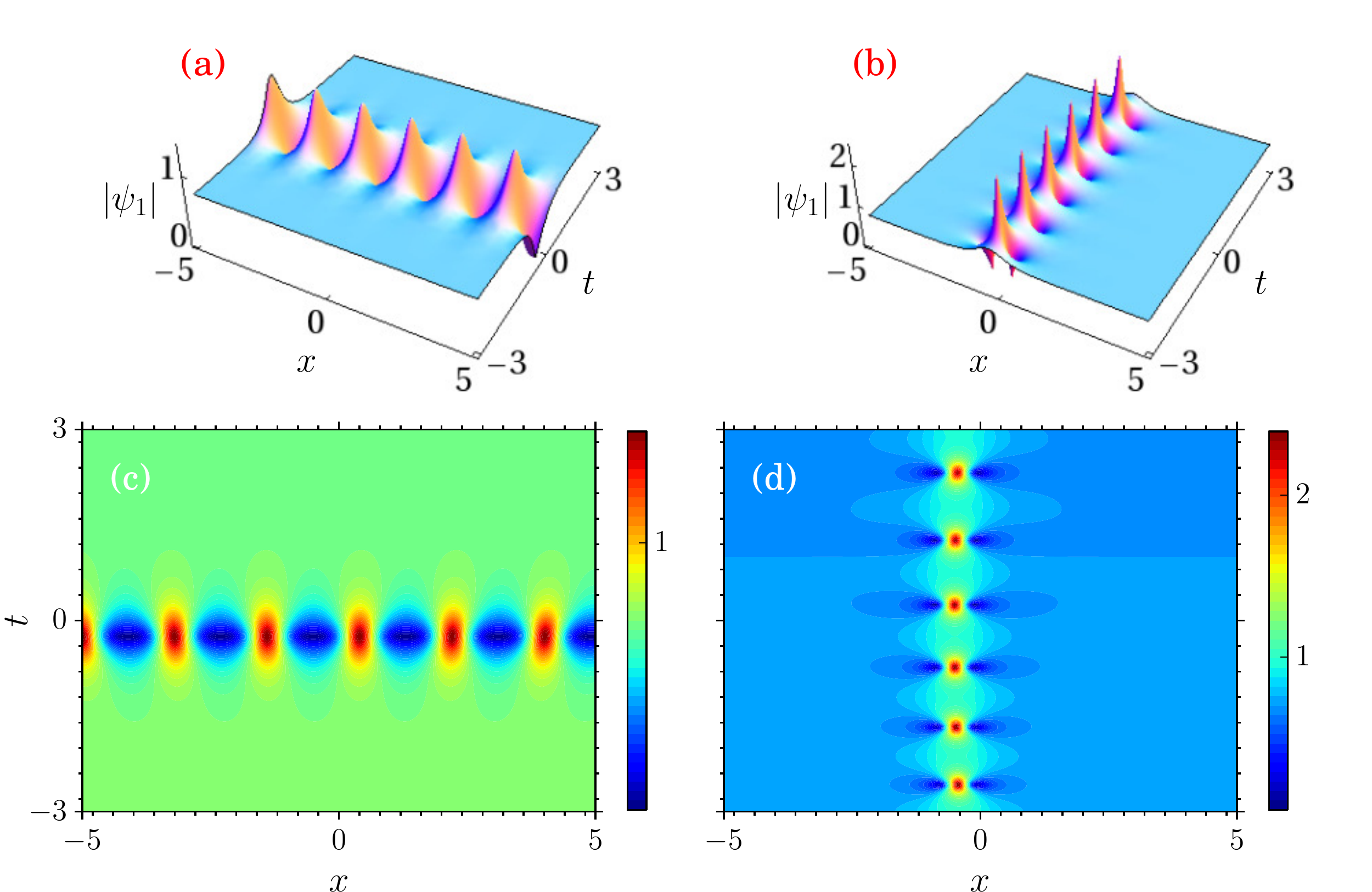}
\end{center}
\caption{(a) AB profile for $l=0.5 I$, (b) Ma breather profile for $l=1.2 I$, (c) and (d) are their corresponding contour plots. The parameters are $\beta_0=2.0$, $\mu_0=2.0$, $\alpha_0=0.01$, $\gamma_0=1$, $\delta_0=0.01$, $\epsilon_0=1$ and $\kappa_0=1$ and $k=0.01$.}
\label{fig1}
\end{figure}

\begin{figure}[!ht]
\begin{center}
\includegraphics[width=0.99\linewidth]{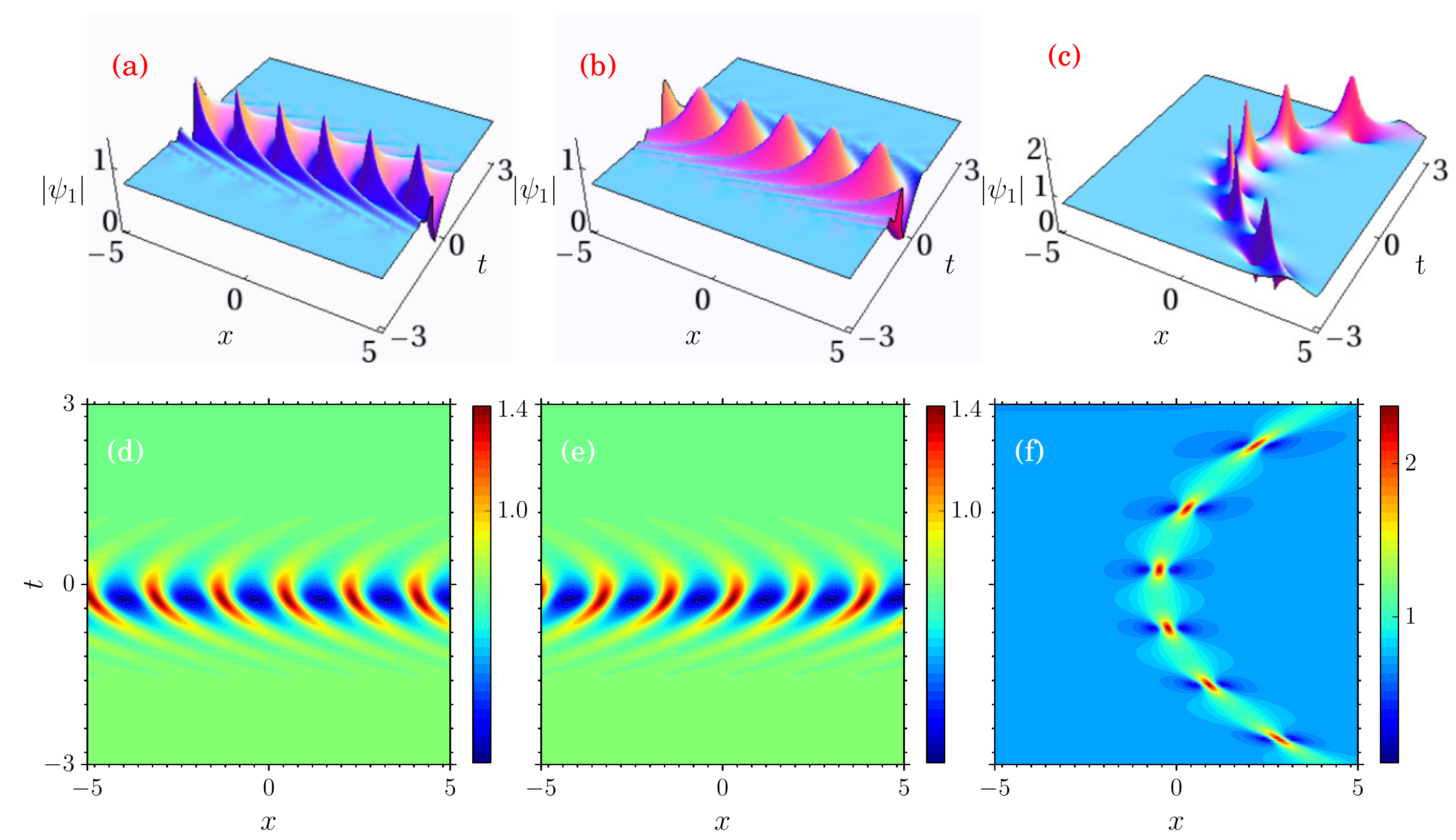}
\end{center}
\caption{(a) AB for $k=2.0$, (b) AB for $k=-2.0$, (c) Ma breather for $k=0.5$, (d), (e) and (f) are their corresponding contour plots. The other parameters are same as in Fig. \ref{fig1}.}
\label{fig1a}
\end{figure}

In Fig. \ref{fig1} we  display the evolution of AB and Ma breather wave solutions of (\ref{a17}) for $k=0.01$.  Fig. \ref{fig1}(a) represents the evolution of an AB for the eigenvalue $l=0.5 i$ and Fig. \ref{fig1}(b) corresponds to Ma breather for the eigenvalue $l=1.2 i$.  The corresponding contour plots are given in the second row.  When we increase the strength of the inhomogeneity parameter ($k$) to $2$, the AB gets stretched in space along the positive $x$ direction which is demonstrated in Fig. \ref{fig1a}(a).  For $k=-2$ the stretching occurs in the reverse direction as shown in Fig. \ref{fig1a}(b).  As for as the Ma breather is concerned, if we $k$ is increase it bends in the plane wave background.  The corresponding structure is given in Fig. \ref{fig1a}(c) for $k=0.5$.  The Ma breather gets curved in the reverse direction for negative value of $k$ which is not displayed here.  The second row in Fig. \ref{fig1a} represents their corresponding contour plots which clearly illustrate how the inhomogeneity parameter influences the breather structures.

We proceed to construct two-breather solutions of (\ref{a17}) and analyze how these solutions get distorted by the inhomogeneity of space.  The two-breather solution of NLS equation is given by \cite{kadz},
\begin{equation}
\label{a18a}
\phi_2(\xi,\tau)=\left[(-1)^j+\frac{G_2(\xi,\tau)+i H_2(\xi,\tau)}{D_2(\xi,\tau)}\right]\exp{(i\tau)},
\end{equation}
where $G_2$, $H_2$, and $D_2$ are given by 
\begin{eqnarray}
G_2 &=& -(k_1^2-k_2^2) \left[\frac{k_1^2\delta_2}{k_2}\cosh(\delta_1\tau_{s1})\cos(k_2\xi_{s2})-\frac{k_2^2\delta_1}{k_1}\cosh(\delta_2\tau_{s2})\cos(k_1 \xi_{s1})  \right. \nonumber \\  && 
\left. - (k_1^2-k_2^2)\cosh(\delta_1\tau_{s1})\cosh(\delta_2\tau_{s2})\right], \nonumber \\
\label{a19}
H_2&=& -2 (k_1^2 - k_2^2) \left[\frac{\delta_1 \delta_2}{k_2} \sinh(\delta_1 \tau_{s1}) -\cos(k_2 \xi_{s2}) - \frac{\delta_1 \delta_2}{k_1}\sinh(\delta_2\tau_{s2}) \cos(k_1 \xi_{s1}) \right. \nonumber \\  &&
\left. -\delta_1 \sinh(\delta_1 \tau_{s1})\cosh(\delta_2 \tau_{s2}) + \delta_2 \sinh(\delta_2 \tau_{s2})\cosh(\delta_1 \tau_{s1})\right],  \\
D_2 &=& 2 (k_1^2 + k_2^2) \frac{\delta_1 \delta_2}{k_1 k_2} \cos(k_1\xi_{s1})\cos(k_2 \xi_{s2}) + 4 \delta_1 \delta_2 (\sin(k_1 \xi_{s1}) \sin(k_2 \xi_{s2}) \nonumber \\ && + \sinh(\delta_1 \tau_{s1}\sinh(\delta_2 \tau_{s2}) - (2 k_1^2 - k_1^2 k_2^2 + 2 k_2^2) \cosh(\delta_1 \tau_{s1})\cosh(\delta_2 \tau_{s2}) \nonumber \\ && - 2 (k_1^2 - k_2^2) \left(\frac{\delta_1}{k_1}\cos(k_1 \xi_{s1})\cosh(\delta_2 \tau_{s2})- \frac{\delta_2}{k_2}\cos(k_2 \xi_{s2}) \cosh(\delta_1 \tau_{s1})\right), \nonumber
\end{eqnarray}
where the modulation frequencies, $k_j=2\sqrt{1+l_j^2}$, $j=1,2$, are described by the (imaginary) eigenvalues $l_j$.  In Eq. (\ref{a19}), $\xi_j, \tau_j$, $j=1,2$, represents the shifted point of origin, $\delta_j(=k_j\sqrt{4-k_j^2}/2)$ is the instability growth rate of each component and $\xi_{sj}=\xi-\xi_j$ and $\tau_{sj}=\tau-\tau_j$ are shifted variables.     
\begin{figure}[!ht]
\begin{center}
\includegraphics[width=0.99\linewidth]{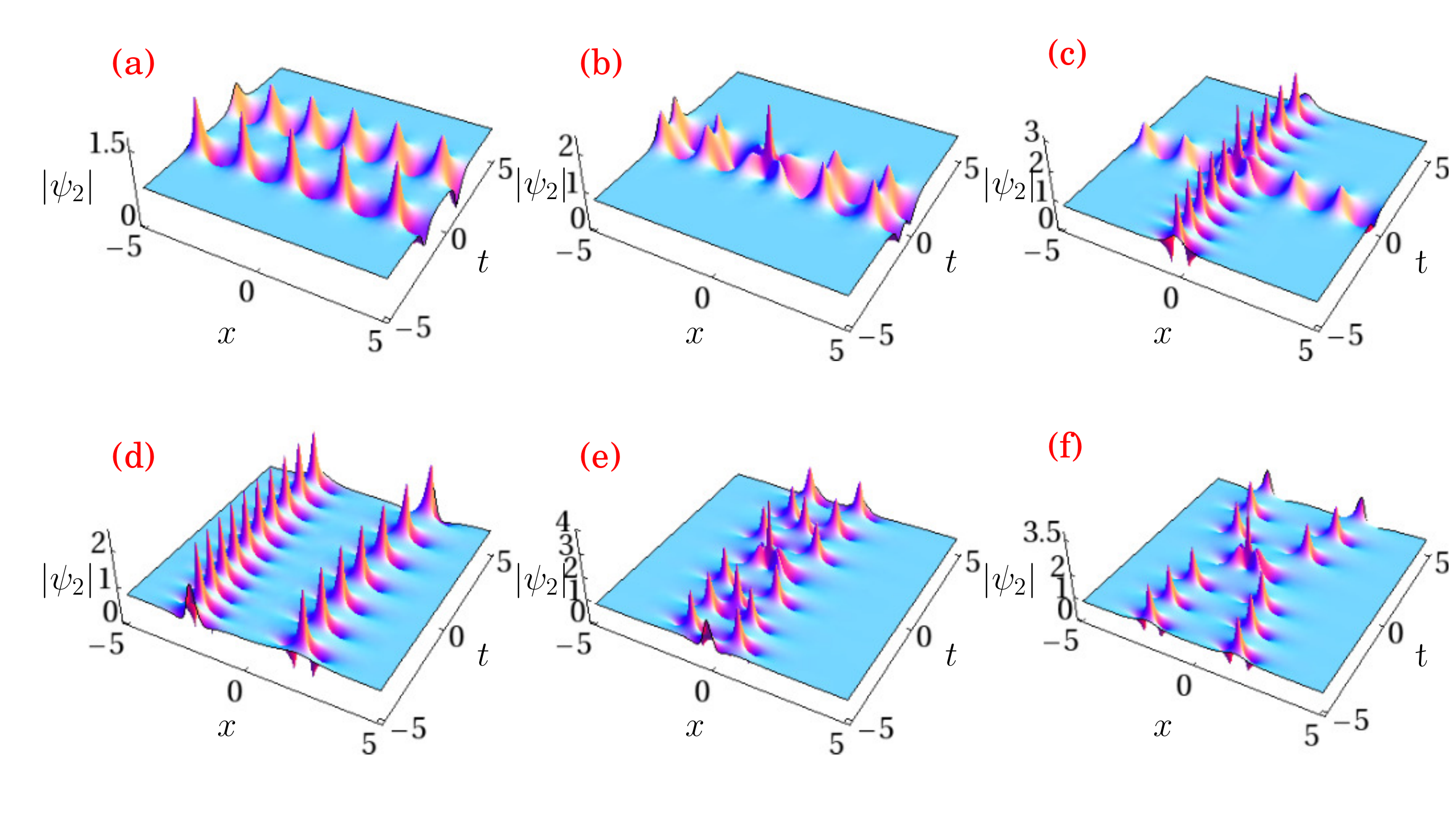}
\end{center}
\caption{(a) Two AB profile for $l_1=0.5i$ and $l_2=0.7i$ with $\tau_1=5$ and $\tau_2=-5$, (b) Two AB without time shifts, (c) the intersection of AB-Ma breathers for $l_1=0.5i$ and $l_2=1.2i$, (d) Two Ma breather for $l_1=1.1i$ and $l_2=1.2i$ with $\xi_1=3$ and $\xi_2=-3$, (e) Two Ma breather without space shifts and (f) Two Ma breather for $l_1=1.1i$ and $l_2=1.11i$. The other parameters are same as in Fig. \ref{fig1}.}
\label{fig2}
\end{figure}

\begin{figure}[!ht]
\begin{center}
\includegraphics[width=0.99\linewidth]{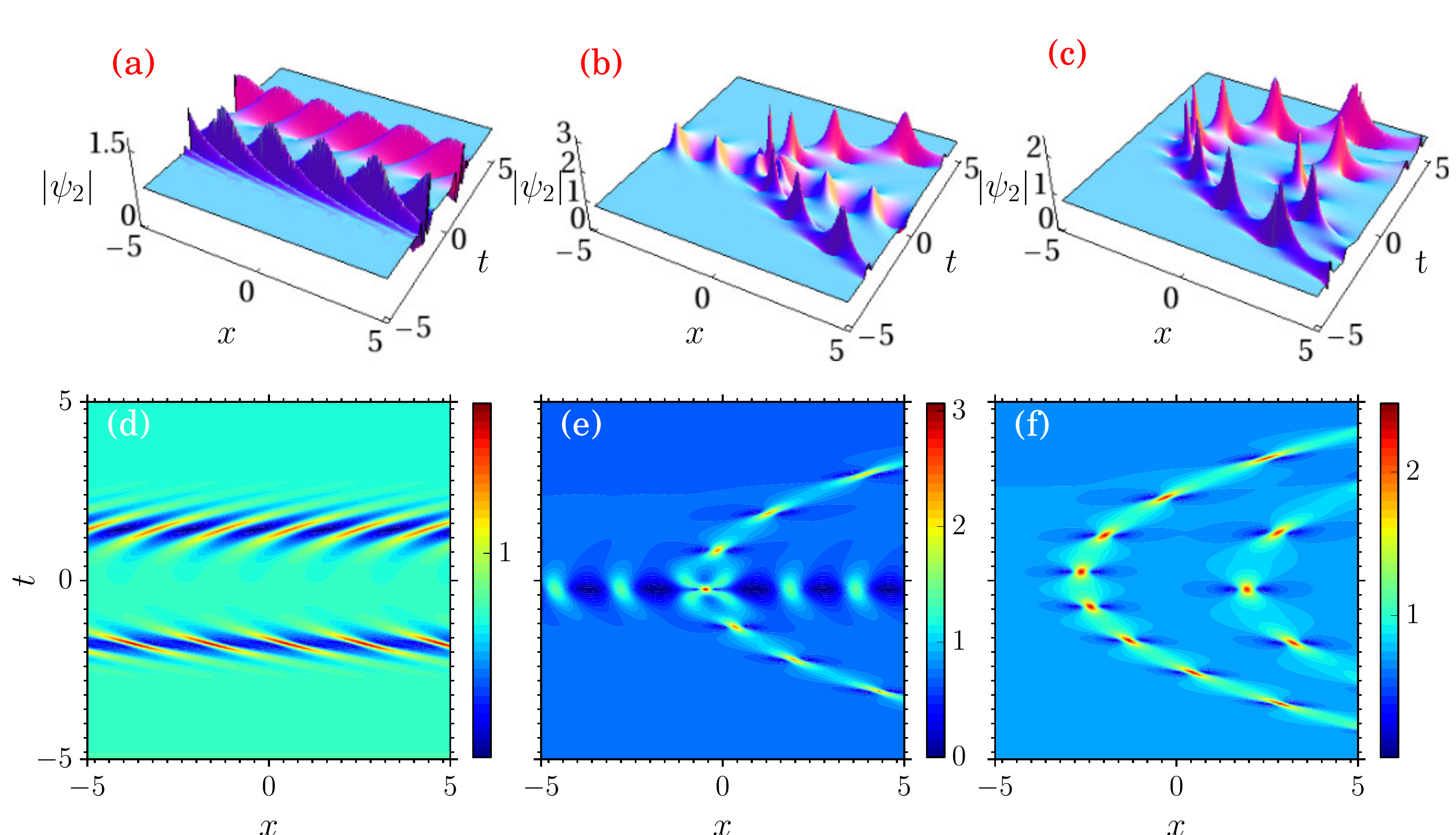}
\end{center}
\caption{(a) Two AB for $k=1.1$, (b) the intersection of AB-Ma breathers for $k=0.5$, (c) two Ma breather for $k=0.5$, (d), (e) and (f) are their corresponding contour plots.}
\label{fig2a}
\end{figure}
With two purely imaginary eigenvalues, $l_j$, $j=1,2$, the solution (\ref{a18a}) is capable of describing a variety of possible second-order breather structures.  The solution includes ABs, Ma solitons and the intersection of AB and Ma breathers in certain combination of eigenvalues.  For example, when both the eigenvalues $Im(l_j)$, $j=1,2$, lie between $0$ and $1$, we obtain the ABs. On the other hand when both of them are greater than one ($Im(l_j)>1$) we obtain the Ma breathers and the mixed possibility, that is one of the eigenvalues is less than one and the other eigenvalue is greater than one, we obtain the intersection of AB and Ma breathers.    

Inserting (\ref{a18a}) in (\ref{a16}) we obtain the general two-breather solution of the INLS Eq. (\ref{a17}).  Fig. \ref{fig2} displays the evolution of two-breather solution of (\ref{a17}) for $k=0.01$ with the imaginary eigenvalues.  To obtain the ABs profile from (\ref{a18a}) we restrict both the eigenvalues $Im(l_1)$ and $Im(l_2)$ to be less than 1 ($l_1=0.5i$ and $l_2=0.7i$).  One AB developing with a time delay after another is shown in Fig. \ref{fig2}(a) and without the time delay is given in Fig. \ref{fig2}(b).  When we change the eigenvalues to $l_1=0.5i$ and $l_2=1.2i$, the AB intersects with Ma breather which is demonstrated in Fig. \ref{fig2}(c).  When both the eigenvalues $Im(l_1)$ and $Im(l_2)$ are greater than 1, say for example $l_1=1.1i$ and $l_2=1.2i$, we obtain two Ma breather solutions from (\ref{a16}).  Similarly the developing of the Ma breather with and without spatial delay is shown in Figs. \ref{fig2}(d) and \ref{fig2}(e) respectively.  We also observe that the distance between the Ma breathers increases when we set both the eigenvalues to be nearly equal, say for example $l_1=1.1i$ and $l_2=1.11i$.  It is shown in Fig. \ref{fig2}(f).  When the strength of the inhomogeneity parameter $k$ is raised to $1.1$, both the ABs are obliquely stretched in $x$ plane which is demonstrated in Fig. \ref{fig2a}(a). When $k=0.5$, the intersection of AB and Ma breather and both the Ma breathers develop a bending structure in the plane wave background.  They are illustrated in Figs. \ref{fig2a}(b) and \ref{fig2a}(c) respectively.  When $k$ is negative, ABs get stretched in the reverse direction.  As we expect AB-Ma/both the Ma breathers develop a bending structure in the negative $x$ direction which is not presented here.  

Recently breathers have been realized experimentally in optical fibers \cite{kibler} and plasmas \cite{bailung}.  The results presented here will also be realized in the experimental context of nonlinear wave propagation in an inhomogeneous plasma.
\section{Characteristics of rogue waves}
Next we move on to investigate the RW solution of (\ref{a17}).  The RW solution can be obtained from the AB/Ma breather solutions as a limiting case \cite{chab:hoff}.  The RW solution has the following basic structure:
\begin{equation}
\label{a22}
\phi_j(\xi,\tau)=\left[(-1)^j+\frac{G_j+i\tau H_j}{D_j}\right]\exp{(i\tau)}, \;\;\; j=1,2,...N,
\end{equation}
where $G_j,H_j$ and $D_j$ are polynomials in $\xi$ and $\tau$. The first-order $(j=1)$ RW solution is given by \cite{akmv:anki}
\begin{equation}
\label{a23}
\phi_1(\xi,\tau)=\left(1-\frac{4(1+2i\tau)}{1+4\xi^2+4\tau^2}\right)\exp{(i\tau)},
\end{equation}
and the second-order $(j=2)$ RW solution is given by  
\begin{equation}
\label{a24}
\phi_2(\xi,\tau)=\left[1+\frac{G_2+i\tau H_2}{D_2}\right]\exp{(i\tau)},
\end{equation}
where
\begin{eqnarray}
G_2&=&\frac{3}{8}-3\xi^2-2\xi^4-9\tau^2-10\tau^4-12\xi^2\tau^2, \nonumber \\
H_2&=&\frac{15}{4}+6\xi^2-4\xi^4-2\tau^2-4\tau^4-8\xi^2\tau^2, \nonumber \\  
D_2 & = &\frac{1}{8}\left(\frac{3}{4}+9\xi^2+4\xi^4+\frac{16}{3}\xi^6+33\tau^2+36\tau^4 
\right. \nonumber \\
& & \left.+\frac{16}{3}\tau^6-24\xi^2\tau^2+16\xi^4\tau^2+16\xi^2\tau^4\right). \nonumber
\end{eqnarray}
\begin{figure}[!ht]
\begin{center}
\includegraphics[width=0.99\linewidth]{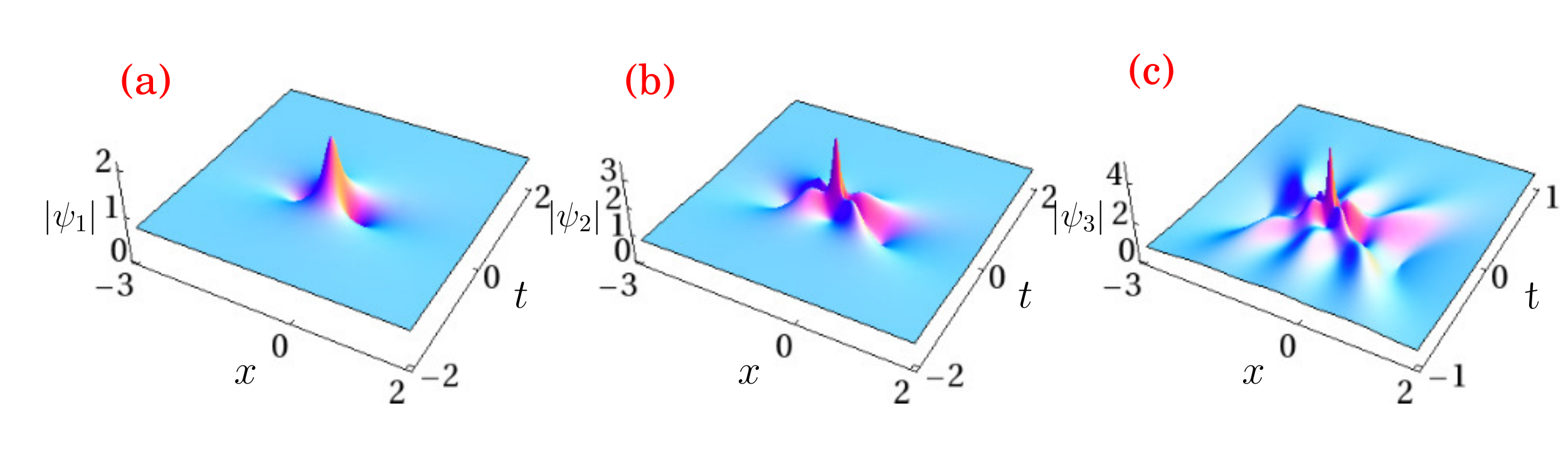}
\end{center}
\caption{(a) First-order RW, (b) second-order RW and (c) third-order RW for $k=0.1$.  The other parameters are same as in Fig. \ref{fig1}.}
\label{fig3}
\end{figure}
When $j=3$ in Eq. (\ref{a22}) we get the third-order $(j=3)$ RW solution \cite{akmv:anki} where $G_3$, $H_3$ and $D_3$ are polynomials in $\xi$ and $\tau$.  Since the explicit expression of third-order RW solution is very lengthy we have not given the underlying expression here and only a graphical analysis of the third-order RW solution is presented.  Substituting (\ref{a23}) and (\ref{a24}) in (\ref{a16}) one can get first and second-order RW solutions of the INLS Eq. (\ref{a17}).  We fix the constants $\mu_0$ and $\beta_0$ to be $2.0$ and display the RW solution of (\ref{a17}) for two different values of the inhomogeneous parameter, say $k=0.1$ and $k=2$, in Figs. \ref{fig3} and \ref{fig3a}.  In Fig. \ref{fig3}, we have given the solution plots of (a) first-, (b) second- and (c) third-order RW solutions for $k=0.1$.  These waves are localized both in space and time thus revealing the characteristic feature of RWs.  When we increase the strength of inhomogeneity parameter $k$ to $2$, we observe that the wave crests get stretched in space and the RW structures get distorted in the plane which is demonstrated in Fig. \ref{fig3a}.  The variations can be seen more clearly in their respective contour plots which are given in the second row in Fig. \ref{fig3a}.  Here also when $k$ is negative the RWs get stretched in the reverse direction.
\begin{figure}[!ht]
\begin{center}
\includegraphics[width=0.99\linewidth]{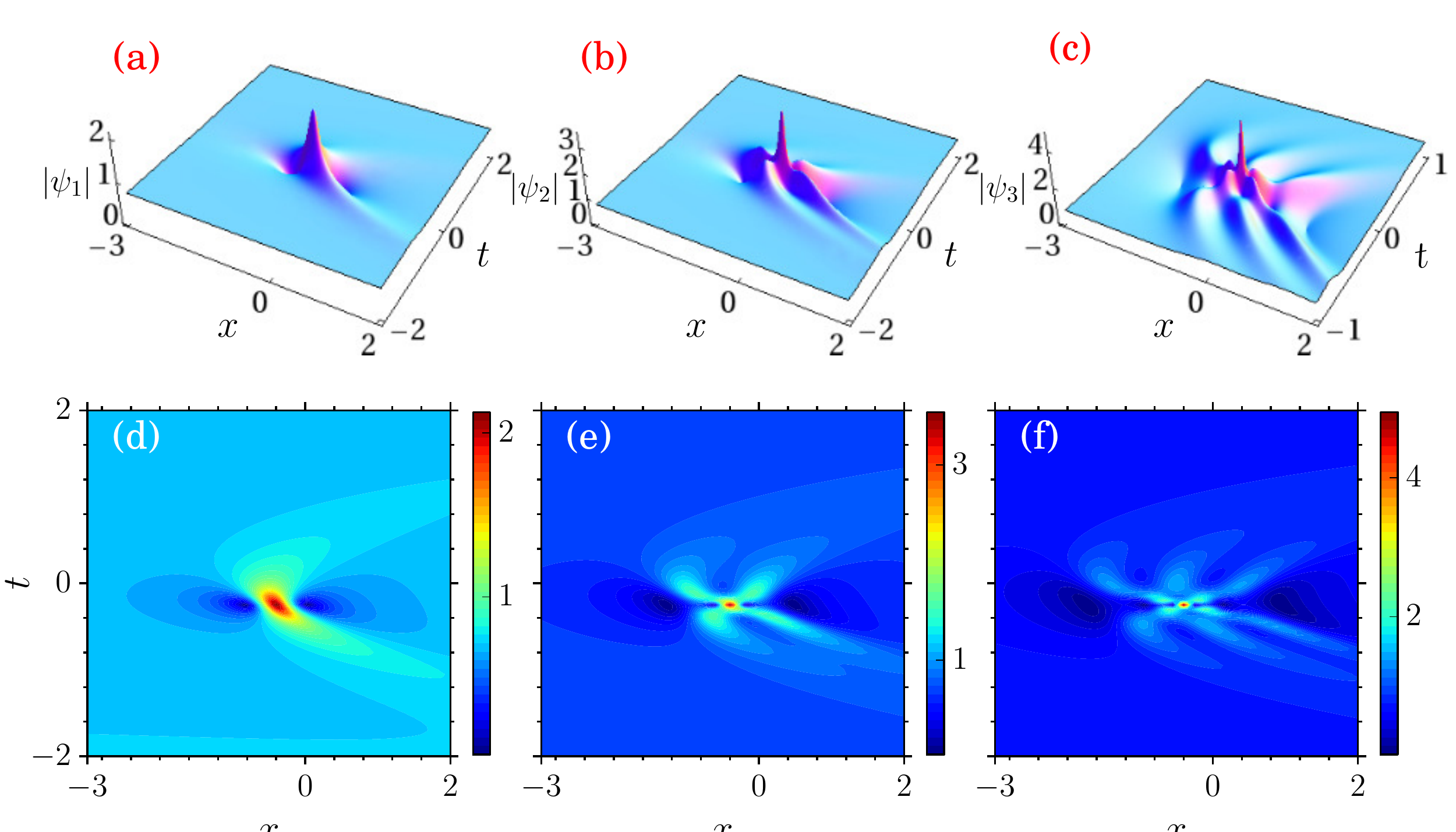}
\end{center}
\caption{(a) First-order RW, (b) second-order RW, (c) third-order RW for $k=2$, (d), (e) and (f) are their corresponding contour plots.}
\label{fig3a}
\end{figure}

Very recently it has been shown that one can also introduce certain free parameters in the RW solutions and by varying these free parameters one can extract certain patterns exhibited by these RWs \cite{ankie}. Motivated by this, in the following, we consider the second- and third-order RW solutions with suitable free parameters and analyze how the RW patterns change with respect to these free parameters for a particular value of inhomogeneity parameter.  To begin with we confine our attention to the second-order RW solution.  In this case, we have the following modified expressions for $G_2$, $H_2$ and $D_2$, that is
\begin{eqnarray}
G_2&=& 12(3-16\xi^4-24\xi^2(4\tau^2+1)-48 l\xi-80\tau^4-72\tau^2-48 m\tau), \nonumber \\
H_2&=& 24 \left[\tau(15-16\xi^4+24\xi^2-48 l\xi-8(1-4\xi^2)\tau^2-16\tau^4)+6m(1-4\tau^2+4\xi^2)\right], \nonumber \\
D_2 & = & 64\xi^6+48\xi^4(4\tau^2+1)+12\xi^2(3-4\tau^2)^2+64\tau^6+432\tau^4+396\tau^2+9 \nonumber \\
& & +48m(18m+\tau(9+4\tau^2-12\xi^2))+48l(18l+\xi(3+12\tau^2-4\xi^2)). \nonumber
\end{eqnarray}
\begin{figure}[!ht]
\begin{center}
\includegraphics[width=0.99\linewidth]{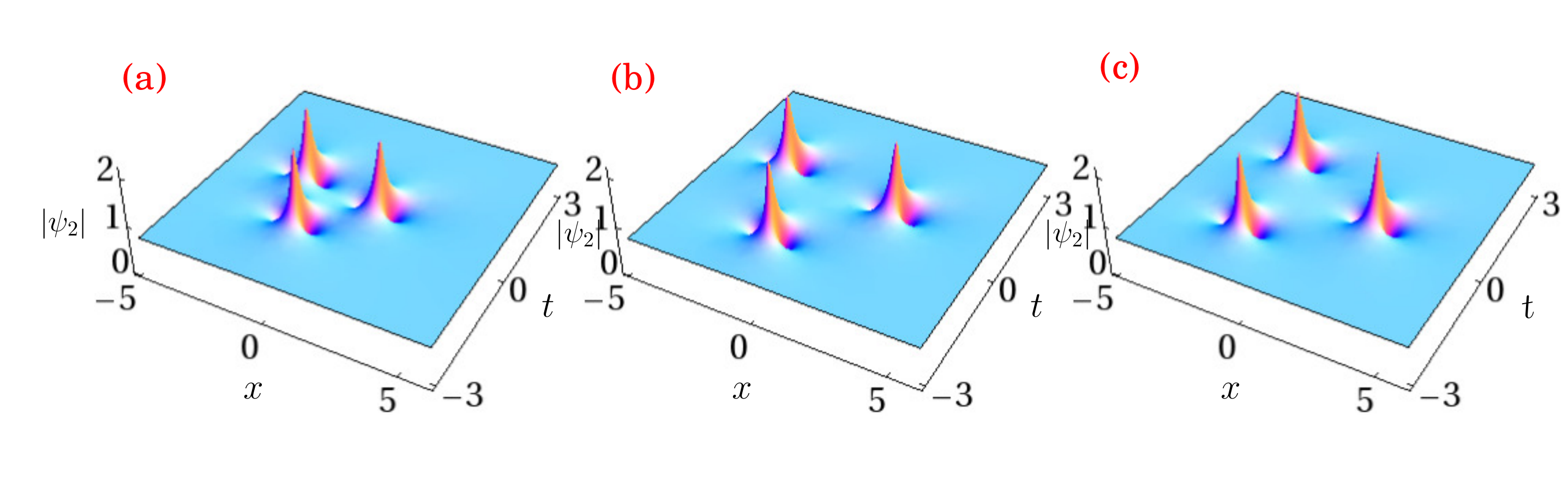}
\end{center}
\caption{RW triplets for $k=0.1$.  Parameters (a) $l=20$ and $m=30$, (b) $l=60$ and $m=100$ and (c) $l=120$ and $m=20$.}
\label{fig4}
\end{figure}
\begin{figure}[!ht]
\begin{center}
\includegraphics[width=0.99\linewidth]{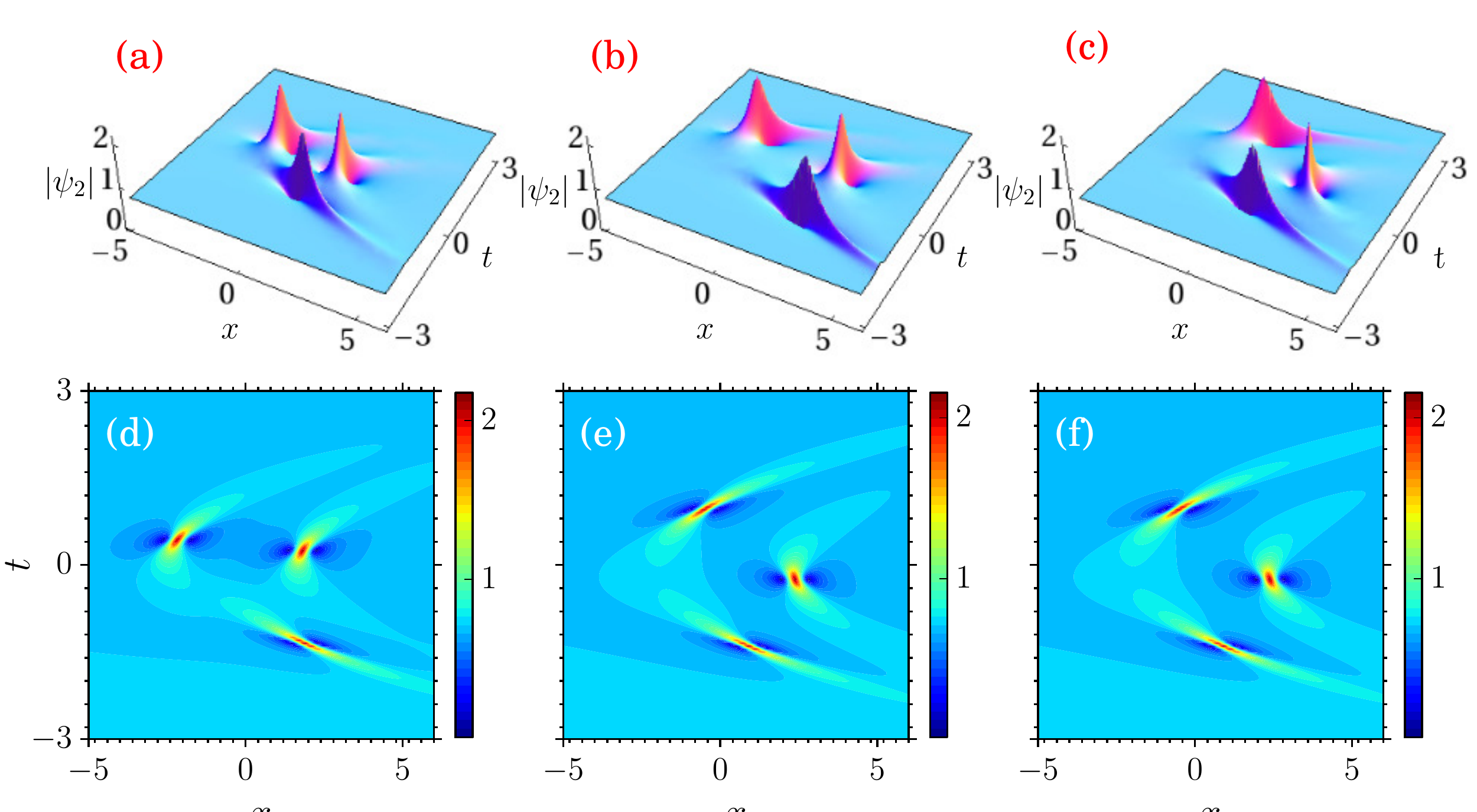}
\end{center}
\caption{(a), (b) and (c) RW triplets when $k=1.5$, (d), (e) and (f) are their corresponding contour plots.}
\label{fig4a}
\end{figure}

This RW solution contains two free parameters, namely $l$ and $m$.  When $l= m=0$, this solution coincides with the one given earlier (vide Eq. (\ref{a24})) which contains one largest crest and four subcrests with two deepest troughs.  When $l$ and $m$ are not equal to 0, the second-order RW splits into three first-order RWs.  These waves emerge in a triangular fashion (a triplet pattern). The parameters $l$ and $m$ describe the relative positions of the first-order RWs in the triplet.  The three first-order RWs form a triangular pattern with $120$ degrees of angular separation between them \cite{ankie}.  We observe this triangular pattern for small values of $l$ and large values of $m$.  On the other hand when $l$ is large and $m$ is small the peaks in the triplet move to new positions in the triangle.  In Fig. \ref{fig4} we display the triplet pattern for $k=0.1$.  The formation of triangular pattern is shown in Fig. \ref{fig4}(a) for $l=20$ and $m=30$.  When we increase the value of $l$ and $m$ to $60$ and $100$ respectively the distance between the peaks in the triplet increases but amplitude of none of the peaks changes which can be seen in Fig. \ref{fig4}(b).  At $l=120$ and $m=20$, the three peaks take new positions which is shown in Fig. \ref{fig4}(c).  When we increase the value of $k$ to $1.5$ the triplet RWs get stretched in space with a curved structure in the plane wave background which is illustrated in Fig. \ref{fig4a}.  When $k$ is negative the triplet RWs get stretched in the reverse direction.
\begin{figure}[!ht]
\begin{center}
\includegraphics[width=0.99\linewidth]{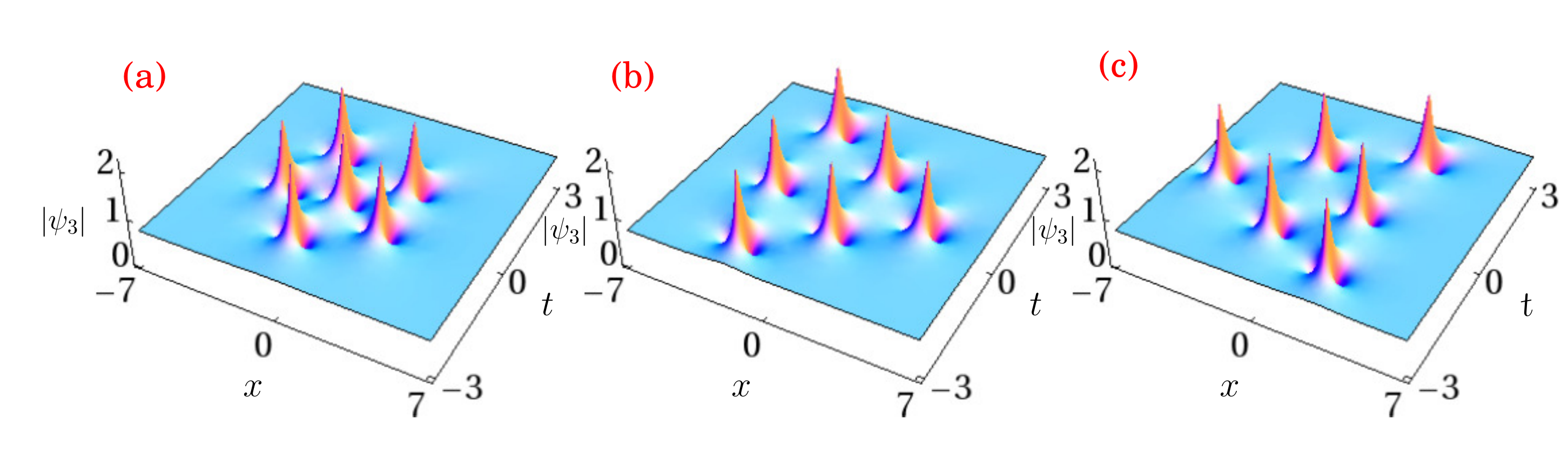}
\end{center}
\caption{RW sextet for $k=0.1$. Parameters (a) $l=10$ and $m=10$, (b) $l=100$ and $m=10$ and (c) $l=-100$ and $m=-10$.} 
\label{fig5}
\end{figure}
\begin{figure}[!ht]
\begin{center}
\includegraphics[width=0.99\linewidth]{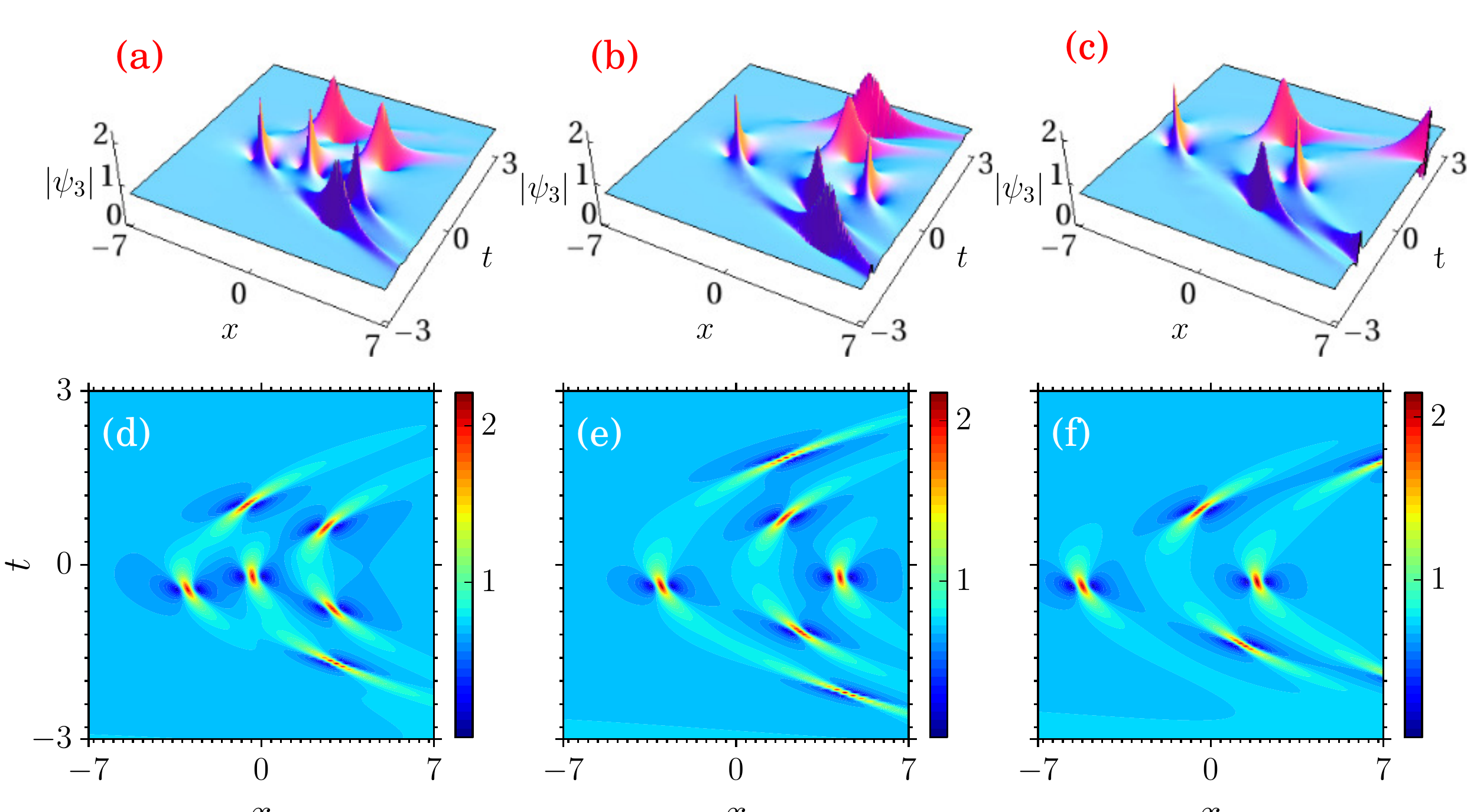}
\end{center}
\caption{(a), (b) and (c) RW sextet when $k=1.5$, (d), (e) and (f) are their corresponding contour plots.} 
\label{fig5a}
\end{figure}

We then move on to investigate the structure of third-order RW solution with four free parameters, namely $l,m,g$ and $e$.  The third-order RW solution with four free parameters is much lengthier than the one without free parameters and so we leave the expression and analyze the results only graphically.  Here also we investigate the solution based on the free parameters.  When $l= m= g= e=0$, we have the classical third-order RW solution which is shown in Fig. \ref{fig3}(c).  It has one largest crest and six subcrests with two deepest troughs.  For non-zero values, the third-order RW splits into six separated first-order RWs.  When we increase the value of free parameters, the six first-order RWs take new positions.  However, the maximum amplitude of each one of the peaks remains the same even in the new orientation.  For small values of $l$ and $m$ and large values of $g$ and $e$ the RWs form a ring structure which contains six peaks.  Figs. \ref{fig5} and \ref{fig5a} display the evolution of third-order RW solution for non-zero values of $l,m,g$ and $e$ for two different values of the inhomogeneous parameter, $k=0.1$ and $k=1.5$ respectively.  When $l= m=10$ and $g= e=500$, we obtain a ring structure with six separated first-order RWs which is shown in Fig. \ref{fig5}(a).  Among the 6 peaks 5 of them assemble on a circle and the sixth one appears in the middle of this circle.  When we increase the value of $l$ to $100$ and keep the other parameters to be $m=10$ and  $g= e=500$ the first-order RWs re-assemble in a triangular form which is shown in Fig. \ref{fig5}(b).  When we fix the values of  $l$ and $m$ to be $-100$ and $-10$ respectively and $g$ = $e=500$ the triangular pattern still persists but the orientation of the triangle changes, which is illustrated in Fig. \ref{fig5}(c). When we increase the value of $k$ to $1.5$, six first-order RWs get stretched in space with a curved structure in the plane wave background which is shown in Fig. \ref{fig5a}.  When $k$ is negative the six first-order RWs get stretched in the reverse direction.

Recent experimental observation of RWs in nonlinear fiber optics \cite{kibler} and in a water tank experiment \cite{chab:hoff} and in plasma \cite{bailung} has open up possibilities to study their characteristics in detail.  Our results, as discussed here, could be useful for controlling highly energetic pulses in plasmas.
\section{Trajectories of inhomogeneous rogue wave}
In this section, we investigate certain characteristics of RW, namely the evolution of its peak, the distance between the valleys', that is width of the RW and the trajectory of the RW analytically.  The trajectory of RW can be described by the motion of the hump and valleys' center location \cite{ling}.  Substituting (\ref{a23}) in (\ref{a16}) one can obtain first-order RW solution of the INLS Eq. (\ref{a17})and we calculate the expression of the motion of its hump's center as
\begin{equation}
\label{xhump}
x_h=-\frac{\epsilon(t)}{\beta(t)},
\end{equation}
and the motions of the two valleys' center is
\begin{figure}[!ht]
\begin{center}
\includegraphics[width=0.99\linewidth]{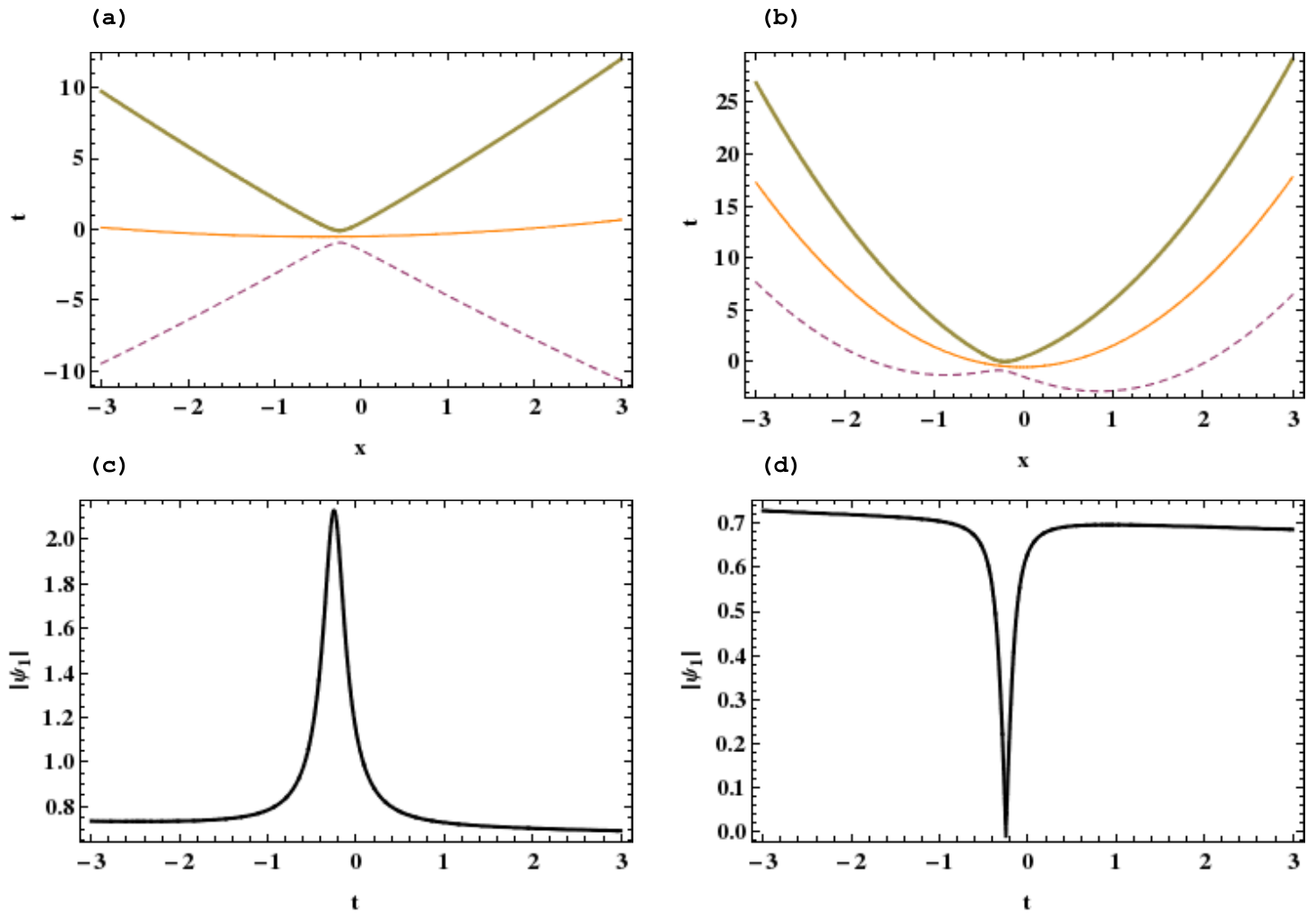}
\end{center}
\caption{The trajectory of the first-order RW when (a) $k=0.1$, (b) $k=2.0$, (c) the evolution of the RW's hump and  (d) the evolution of the RW's valleys. The other parameters are same as in Fig. \ref{fig1}.} 
\label{fig6}
\end{figure} 
\begin{equation}
\label{xval}
x_v=\frac{\pm \left(\sqrt{3\beta(t)^2(1+4 \gamma(t)^2)}\right) - 2 \beta(t) \epsilon(t)}{2 \beta(t)^2},
\end{equation}
where $x_h$ and $x_v$ denote the trajectory of hump and valleys of RW respectively.  The trajectory of the RW's hump is shown by orange line and the trajectories of the two valleys look like an "X" shape, as shown by the thick and dashed lines in Fig. \ref{fig6}(a).  The corresponding evolution plot is shown in Fig. \ref{fig3}(a).  When the inhomogeneity parameter $k=2$, the trajectory of the RW is depicted in Fig. \ref{fig6}(b) and its corresponding evolution diagram is shown in Fig. \ref{fig3a}(a).  Furthermore, we can define the width of the RW as the distance between the two valleys' centers \cite{ling}.  Its evolution is 
\begin{equation}
\label{width}
W(t)_{rw}=\frac{\sqrt{3\beta(t)^2(1+\gamma(t)^2)}}{\beta(t)^2}.
\end{equation}
Substituting (\ref{xhump}) and (\ref{xval}) into the expression of (\ref{a16}), we can calculate the expressions for the evolution of the RW's hump and valleys as shown in Figs. \ref{fig6}(c) and (d) respectively.

\section{Conclusion}
In this work, we have considered a variable coefficient NLS equation with an external linear potential that describes the nonlinear wave propagation in an inhomogeneous plasma/medium.  We have constructed several localized solutions for the INLS equation, including Ma breather, AB, two-breathers, first-, second- and third-order RW solutions by mapping it to the NLS equation.  Our aim was to investigate in-detail the impact of inhomogeneity on all these rational solutions.  When the inhomogeneity parameter $k$ is small, say $0.01$, we have noticed that the breather and RW solutions retain their shapes.  For $k=1.5$, the shape of AB gets stretched and the Ma breather bends in the plane wave background.  For negative $k$ values the same effects occur but in the reverse direction.  When we increase the strength of inhomogeneity parameter in two-breather solutions, both the ABs obliquely stretches in space whereas the two Ma breathers and the intersection of AB and Ma breathers get curved in the plane wave background.  Similar effects were also observed for the RW solutions.  For example, when we increase the value of $k$ in the INLS equation, the RW solution get stretched in space or curved in the plane wave background.  We have constructed the second- and third-order RW solutions with certain free parameters and also analyzed how they are modified by the inhomogeneity parameter.  We have derived the second-order RW solution with two free parameters.  For small values of these free parameters we have the triangular pattern of three separated first-order RWs and when we increase the value of free parameters the distance between the peaks in the triplet increases and these peaks occupy a different position.  Finally, we have derived the third-order RW solution with four free parameters.  By varying these free parameters we have obtained two different patterns of six first-order RWs, namely (i) the ring structure and (ii) the triangular structure.  We have also investigated the impact of inhomogeneity on these solutions as well.   The nonlinear structures, as reported here, may be useful for controlling plasmonic energy along the plasma surface.  Our results provide the many possibilities to manipulate RWs both theoretically as well as experimentally in their relative fields such as fluids, nonlinear optics and Bose-Einstein condensates.

\section*{Acknowledgements}
KM thanks the University Grants Commission (UGC-RFSMS), Government of India, for providing a Research Fellowship.  The work of MS forms part of a research project sponsored by NBHM, Government of India.


\section*{References}
\begin{thebibliography}{90} 

\bibitem{budd}
K.G. Budden, Radio Wave in the Ionosphere, Cambridge University Press, London, 1961.

\bibitem{osbore}
A.R. Osborne, Nonlinear Ocean Waves, Academic Press, New York, 2009.

\bibitem{hase}   
A. Hasegawa, Opt. Lett. {\bf 5} (1980) 416-417.

\bibitem{tewari}
D.P. Tewari, R.R. Sharma, J. Phys. D: Appl. Phys. {\bf 12} (1979) 1019.

\bibitem{chen:liu}
H.H. Chen, C.S. Liu, Phys. Rev. Lett. {\bf 37} (1976) 693; H.H. Chen, C.S. Liu, Phys. Fluids {\bf 21} (1978) 377.

\bibitem{bala}
R. Balakrishnan, Phys. Rev. A {\bf 32} (1985) 1144.

\bibitem{serkin}
V.N. Serkin, A. Hasegawa, T.L. Belyaeva, Phys. Rev. Lett. {\bf 98} (2007) 074102.

\bibitem{atre:pani}
R. Atre, R.K. Panigrahi, G.S. Agarwal, Phys. Rev. E {\bf 73} (2006) 056611.

\bibitem{solli}
D.R. Solli, C. Ropers, P. Koonath, B. Jalali, Nature {\bf 450} (2007) 1054.

\bibitem{yan}
Z.Y. Yan, Phy. Lett. A {\bf 374} (2010) 672.

\bibitem{zhong}
W.P. Zhong, R.H. Xie, M. Belic, N. Petrovic, G. Chen, Phys. Rev. A {\bf 78} (2008) 023821.

\bibitem{belm}
J. Belmonte-Beitia, V.M. Perez-Garcia, V. Vekslerchik, V.V. Konotop, Phys. Rev. Lett. {\bf 100} (2008) 164102.

\bibitem{haseg}
A. Hasegawa, M. Matsumoto, Optical Solitons in Fibers, Springer-Verlag, Berlin, 2003.

\bibitem{dai}
C.Q. Dai, C.L. Zheng, H.P. Zhu, Eur. Phys. J. D {\bf 66}, (2012) 112.

\bibitem{yang}
Z. Yang, W. Zhong, M.R. Belie, Phys. Scr. {\bf 86}, (2012) 015402.

\bibitem{xfwu}
X.F. Wu, G.S. Hua, Z.Y. Ma, Commun. Nonlinear. Sci. Numer. Simulat. {\bf 18} (2013) 3325.

\bibitem{wang}
Y.Y. Wang, J.S. He, Y.S. Li, Commun. Theor. Phys. {\bf 56} (2011) 995.

\bibitem{khaw}
U. Al Khawaja, M. Taki, Phys. Lett. A {\bf 377} (2013) 2944.

\bibitem{shuk}
W.M. Moslem, P.K. Shukla, B. Eliasson, Euro. Phys. Lett. {\bf 96} (2011) 25002.

\bibitem{mande}
D. Mandelik, H.S. Eisenberg, Y. Silberberg, R. Morandotti, J.S Aitchison, Phys. Rev. Lett. {\bf 90} (2003) 253902.

\bibitem{karif}
C. Kharif, E. Pelinovsky, Eur. J. Mech. B/Fluids {\bf 22} (2003) 603.

\bibitem{kharif}
C. Kharif, E. Pelinovsky, A. Slungyaev, Rogue waves in the Ocean, Springer, Heidelberg, 2009.

\bibitem{osborn}
A.R. Osborne, M. Onorato, M. Serio, Phys. Lett. A {\bf 275} (2000) 386.

\bibitem{blud:kono}
Y.V. Bludov, V.V. Konotop, N. Akhmediv, Phys. Rev. A {\bf 80} (2009) 033610.

\bibitem{kibler}
B. Kibler, J. Fatome, C. Finot, G. Millot, F. Dias, G. Genty, N. Akhmediev, J. M. Dudley, Nat. Phys. {\bf 6} (2010) 790.

\bibitem{chab:hoff}
A. Chabchoub, N.P. Hoffmann, N. Akhmediev, Phys. Rev. Lett. {\bf 106} (2011) 204502.

\bibitem{bailung}
H. Bailung, S.K. Sharma, Y. Nakamura, Phys. Rev. Lett. {\bf 107}, (2011) 255005.

\bibitem{shat}
M. Shats, H. Punzmann, H. Xia, Phys. Rev. Lett. {\bf 104}, (2010) 104503.

\bibitem{ya}
Z.Y. Yan, Commun. Theor. Phys. {\bf 54}, (2010) 947.

\bibitem{mosl}
W.M. Moslem, P.K. Shukla, B. Eliasson, Europhys. Lett. {\bf 96} (2011) 25002.

\bibitem{pere}
D.H. Pergrine, J. Austral. Math. Soc. Ser. B {\bf 25} (1983) 16. 

\bibitem{akmv:anki}
N. Akhmediev, A. Ankiewicz, J.M. Soto-Crespo, Phys. Rev. E {\bf 80} (2009) 026601.

\bibitem{ma}
Y.C. Ma, Stud. Appl. Math. {\bf 60} (1979) 43.

\bibitem{eleon}
N. Akhmediev, V.M. Eleonskii, N.E. Kulagin, Zh. Eksperimentalnoii i Teoreticheskoii Fiziki {\bf 89} (1985) 1542  [Sov. Phys. JETP {\bf 62},  (1985) 894]

\bibitem{kadz}
D.J. Kedziora, A. Ankiewicz, N. Akhmediev, Phys. Rev. E {\bf 85}, (2012) 066601.

\bibitem{susl}
S.K. Suslov, Proc. Ame. Mat. Soc. {\bf 140} (2012) 3067.

\bibitem{pono}
S.A. Ponomarenko, G.P. Agrawal, Phys. Rev. Lett. {\bf 97} (2006) 013901.

\bibitem{ankie}
A. Ankiewicz, D.J. Kedziora, N. Akhmediev, Phy. Lett. A {\bf 375} (2011) 2782-2785.

\bibitem{ling}
L. Ling, L.C. Zhao, Phys. Rev. E {\bf 88} (2013) 043201.

\end{thebibliography}
\end{document}